\input harvmac

\def\p{\partial}

\def\b0{\bar{0}}
\def\b4{\bar{4}}

\Title{EFI-97-46}{\vbox{\centerline{Matrix Schwarzschild Black Holes in
Large N limit}}}
\vskip20pt

\centerline{Miao Li}
\bigskip
\centerline{\it Enrico Fermi Institute}
\centerline{\it University of Chicago}
\centerline{\it 5640 Ellis Avenue, Chicago, IL 60637, USA} 
\centerline{\tt mli@curie.uchicago.edu}

\bigskip
\centerline{\it }
\centerline{\it }
\centerline{\it } 
\bigskip

Based on a gas picture of D0-brane partons, 
it is shown that the entropy, as well as the geometric size of an infinitely
boosted Schwarzschild black hole, can be accounted for in matrix theory
by interactions involving spins, or interactions involving
more than two bodies simultaneously.

\Date{October 1997} 

\nref\bfss{T. Banks, W. Fischler, S. H. Shenker and L. Susskind,
hep-th/9610043.}
\nref\lm{M. Li and E. Martinec, hep-th/9703211; hep-th/9704134.}
\nref\dvv{D. Dijkgraaf, E. Verlinde and H. Verlinde, hep-th/9704018.}
\nref\halyo{E. Halyo, hep-th/9705107.}
\nref\bfks{T. Banks, W. Fischler, I. Klebanov and L. Susskind,
hep-th/9709091; I. Klebanov and L. Susskind, hep-th/9709108;
E. Halyo, hep-th/9709225.}
\nref\hm{G. Horowitz and E. Martinec, hep-th/9710217.}
\nref\dkps{M.R. Douglas, D. Kabat, P. Pouliot and S.H. Shenker,
hep-th/9608024.}
\nref\jh{J. Harvey, hep-th/9706039.}
\nref\aclm{H. Awata, S. Chaudhuri, M. Li and D. Minic, hep-th/9706083.}
\nref\kraus{P. Kraus, hep-th/9709199.}
\nref\dr{M. Dine and A. Rajaraman, hep-th/9710174.}
\nref\ms{J. Maldacena and L. Susskind, hep-th/9604042.}
\nref\seiberg{N. Seiberg, hep-th/9705221.}
\nref\ss{A. Sen, hep-th/9709220; N. Seiberg, hep-th/9710009.}

\newsec{\it Introduction}

One of the more interesting applications of matrix theory \bfss\ is
to the study of quantum properties of black holes. Some first steps in
this direction have been taken in \refs{\lm, \dvv,\halyo}. More recently,
discussions on Schwarzschild black holes at a special kinematic
point are presented in \bfks. However, as pointed out in \hm, the work
of \bfks\ represents certain understanding of the black string side
right before the black string collapses in the longitudinal direction
to form a black hole.

To truly understand the hole regime in the infinite momentum frame (IMF),
one has to look at the limit $N>>S$, where $N$ is the number of partons
and $S$ is the entropy of the hole. For a given rest mass $M$, only in
this limit could one hope that the entropy becomes independent of how
much one boosts the black hole. And it appears, as also emphasized 
in \hm, that only the zero-modes of the relevant large N Yang-Mills theory
are significant in forming the cluster, as the temperature is so low
in the IMF. We shall present a plausible picture for understanding
Schwarzschild black holes above four dimensions along this line.

Compactifying M-theory on $T^d$, we are left with $D=11-d$ dimensional
uncompactified spacetime. For simplicity, $T^d$ is assumed to be a 
rectangular torus with equal circumference $L$.
To formulate the matrix theory, we need to
compactify one longitudinal dimension too with radius $R$, and eventually
take the limit $R\rightarrow\infty$ after relevant calculations are done.
We start with the standard formulas concerning a Schwarzschild black hole
\eqn\bh{R_s=(G_DM)^{{1\over D-3}}, \quad S=R_s^{D-2}/G_D=M^{{D-2\over
D-3}}
G_D^{{1\over D-3}},}
where $G_D=l_p^9/L^d$ is the Newton constant in $D$ dimensions. These two
equations are the ones we wish to derive in the matrix theory context
up to numerical coefficients. The on-shell relation in the IMF
$$E_{LC}= {M^2 R\over N}$$
together with the thermodynamics relation $E_{LC}=ST$ results in
\eqn\supp{S={M^2 R\over NT}.}
As long as $D>4$, this relation and the second equation in \bh\ combine
to yield
\eqn\mass{M=(NT/R)^{{D-3\over D-4}}G_D^{{1\over D-4}},}
and
\eqn\entropy{S=(NT/R)^{{D-2\over D-4}}G_D^{{2\over D-4}}.}
This is the equation of state predicted by the Bekenstein-Hawking entropy
formula. Furthermore, we have
\eqn\radius{R_s=(NTG_D/R)^{{1\over D-4}}.}

In the next few sections we aim to give an explanation of 
eqs.\entropy\ and \radius.

\newsec{\it The hole limit} 

We are interested in the hole limit
where $R>>R_s$. With the help of \radius\ this is
\eqn\lim{R^{D-4}>>{NTG_D\over R}.}
For the IMF physics to work effectively, $N/R>>M$, this together with
\mass\ gives
\eqn\limi{R>>NT^{D-3}G_D.}
The two conditions \lim\ \limi\ imply $N>>RM>>R_sM=S$, that is, we are 
far from the point $N=S$ considered in \bfks\ and \hm. Finally, to trust
thermodynamics, $S>>1$ which implies
\eqn\limit{R<<NTG_D^{{2\over D-2}}.}
Combination of three inequalities \lim\ \limi\ \limit\ 
for $R$ yields conditions on the temperature
\eqn\temp{NT>>G_{D}^{-{1\over D-2}}>>T,}
where $G_{D}^{-{1\over D-2}}$ is just the Planck mass in $D$ dimensions.

As we shall show toward the end of this paper, the temperature is so
low that it is difficult, if not impossible,  to do thermodynamics on the 
dual torus
$\tilde{T}^d$ on which the Supersymmetric Yang-Mills is defined. This 
indicates that, the physics is dominated by the zero-modes representing
D0-brane dynamics, and the desired formulas \entropy\ and \radius\
must be derived within the picture of D0-brane gas. Later we shall show
that indeed the Born-Oppenheimer approximation is good in the large N
limit provided $D>4$, and this is just the condition for the validity of
\entropy\ and \radius.

\newsec{\it The interacting gas picture} 

As argued convincingly 
in \hm, the hole limit must be dominated by the zero-mode dynamics
of the underlying large N theory. Our first piece of firm evidence
for this is the estimate of the geometric radius $R_s$, given knowledge
about the entropy as in \entropy. We adapt a calculation in the first
paper of \bfks. Within the gas picture, the black hole is thought of
as a long-lived bound state of partons, and the virial theorem is 
applicable here. The kinetic energy of partons and the total energy 
are of the same order, thus
\eqn\virial{Nm<v^2>=TS=T(NT/R)^{{D-2\over D-4}}G_D^{{2\over D-4}}.}
Now the mean velocity is determined by the size of the bound state $R_s$
and the typical frequency which we take to be $T$, so $v\sim TR_s$.
Substituting this into the above equation we obtain a relation between
$R_s$ and $S$. Indeed, given the R.H.S. of \virial\ we determine
\eqn\rad{R_s=(NTG_D/R)^{{1\over D-4}},}
precisely the desired result \radius. The fact that such a simple estimate
gives us the correct scaling strongly suggests that the interacting 
gas picture is a good one. The virial theorem thus reduces the two
independent unknowns $R_s$ and $S$ to only one. If there is a way
to determine one of them, then both are determined correctly.

Our strategy for determining $R_s$ is the following. We first postulate
some relevant forms of interaction energy, and use the desired result
for $R_s$ to determine them. We then argue that these interactions 
exist in matrix theory, and assuming certain correlations among spin
and orbital motion, these are dominant interactions, thus justifying
the calculation.

For simplicity, we assume the dominant interaction depend on
the mean velocity and the mean separation between partons in a power
law fashion. Spins will be important, and for our purpose we can always
choose a proper normalization such that they do not figure in for
the moment. The total interaction will also depend on $N$, after summing
up over partons. Let $V_l=C_{N,l}v^{l+1}/r^n$ be the total interaction
energy. The meaning of $l$ will become clear momentarily. By the virial
theorem,
\eqn\viria{Nmv^2\sim C_{N,l}v^{l+1}/R_s^n.}
Plugging $v\sim TR_s$ into \viria, and reading off the dependence of $R_s$
on $T$, we determine $n$
\eqn\num{n=(l-1)(D-3),}
in order to match on to \radius. To match the whole formula, then
$C_{N,l}=(N/R)^lG_D^{l-1}$, so the total interaction energy is
\eqn\poten{V_l=\left({N\over R}\right)^lG_D^{l-1}{v^{l+1}\over 
r^{(l-1)(D-3)}}.}

Now, the dependence of $V_l$ on $N$ suggests to us that the origin of
$V_l$ is a $l$-body interaction, since the total number of $l$-tuples
is of the order $N^l$. The first choice is $l=2$, the power in $v$ goes
as $v^3$ and implies that some dependence on spin is needed. The second
choice is $l=3$, a 3-body interaction. As we shall show, for all $l$, the
interaction as given in \poten\ is possible in matrix theory. Before doing
that, we want to compare the contribution of $V_l$ to the well-known
two-body interaction $U_2=N^2(G_D/R^3)v^4/r^{D-4}$. Since all $V_l$ are of
the same order, it is sufficient to do this for $V_2$:
\eqn\compare{{V_2\over U_2}={R\over vR_s}={N\over R^{D-2}_s/G_D}={N\over S}
>>1.}
where we used the formula for $R_s$ to express $T$ in terms of other 
quantities.
Indeed, all the interaction forms \poten\ dominate over the standard
velocity dependent two-body interaction in the regime where we have
an infinitely boosted black hole.

Next we turn to the issue whether these desired interactions can actually
arise in matrix theory. The $l$-body interaction \poten\ is to be calculated
as a scattering amplitude in which $l$ partons scatter into $l$ partons.
For a given Feynman diagram in the matrix quantum mechanics, typically
one need to insert an operator for each outleg. Now there are $2l$ outlegs,
and only $l+1$ velocity factors, apparently we need a factor 
$(\psi^2)^{l-1}$ to make up all the insertions, where $\psi$ is
the fermion of 16 components. Thus, schematically,
the $l$-body interaction is
\eqn\intera{u_l=({G_D^{l-1}\over R^l})v^{l+1}(\psi^2)^{l-1}/r^{(l-1)
(D-3)}.}
Note that, for $l=2$, this two-body spin dependent force is computed 
in \jh\ and discussed in \aclm, whose existence is therefore confirmed.
For a recent calculation in matrix theory, see \kraus.

To see whether other $l$-body interactions can be derived in matrix 
theory, we first concentrate on the case $D=11$, and then argue for
general $D$. To this end, we need to write down the matrix action 
schematically
\eqn\matrix{S={1\over \hbar}\int dt\Tr\left( (\p_t X)^2
+[X,X]^2+\psi\p_t\psi +\psi [X,\psi]\right),}
where $\hbar=R^3/l_p^6$. In putting the action into the form \matrix,
we have rescaled $X\rightarrow  (l_p^3/R)X$, so that $X$ in \matrix\
has the dimension $[X]=L^{-1}$. Similarly, $[\psi]=L^{-3/2}$. 
Now, $[v^{l+1}(\psi^2)^{l-1}/r^{8(l-1)}]=L^{3l-7}$, where we have put 
$D=11$. To obtain a term with dimension of energy, a factor $\hbar^{l-2}$
is to be inserted. This means that the $l$-body effect is of 
$l-1$ loops. For $l=2$, this agrees with the analysis of \refs{\jh,
\aclm}. For $l=3$, this is a two-loop effect.
Next, we want to check whether the dimensional coefficient comes
out correctly. Rescaling $r$ back, we have a factor $(l_p^3/R)^{7l-9}$,
this together with $\hbar^{l-2}=(R^3/l_p^6)^{l-2}$ gives
$R^{-l}(l_p^{15}/ R^3)^{l-1}$. Finally, $\psi^2$ as a spin factor scales
as $\hbar$, so $(\psi^2)^{l-1}$ contributes a factor $\hbar^{l-1}
=(R^3/l_p^6)^{l-1}$. This combined with the previous factor we obtained
gives $R^{-l}l_p^{9(l-1)}=R^{-l}G_{11}^{l-1}$, the right combination
appearing in \intera.

Demanding that the above result directly generalizes to $D$ dimensions
requires the distance dependent part assume a special form. For the
$l$-body interaction, one has to sum over periodic images of $l-1$ partons
on $T^d$. Pick $x_1$ out and assume that the dependence of the
separations is $\prod_{i=2}^l|x_1-x_i|^{-8}$. Now summing over all the images
of $x_i$ ($i\ge 2$) one obtains $L^{-d(l-1)}\prod_{i=2}^l|x_1-x_i|^{-(D-3)}$. 
The factor $L^{-d(l-1)}$ is precisely the one needed to yield
$G_D^{l-1}$ from $G_{11}^{l-1}$. For $l=3$, we note that a similar 3-body
interaction is discussed in a recent paper \dr. Our interaction
$v^4(\psi^2)^2/r^{16}$ as in $D=11$ is a super-partner of 
$v^6/r^{14}$. The term ruled out in \dr\ however is not the same term
as ours, since the separation dependent part of that term depends on
all $r_{ij}=|x_i-x_j|$, and there it is assumed $r_{12}\sim r_{13}=R
>>r_{23}=r$. 

Finally, we need to justify the Born-Oppenheimer approximation. Following
\dkps, for a typical velocity $v$, there is a characteristic size called
the stadium size, below which the the Born-Oppenheimer approximation
breaks down. In $D$ dimensions, the stadium size is just $\sqrt{v}
\sqrt{l_p^3/R}$. Now $v\sim TR_s$, so the stadium size goes like
$\sqrt{TR_s}\sqrt{l_p^3/R}$.  For a given cut-off $R$ (no matter how large 
it is) and a given fixed horizon size $R_s$, the temperature scales
as $1/N$ according to \radius\ in the large N limit. Thus the stadium
size scales as $1/\sqrt{N}$. The mean separation between the nearest
two partons is $R_s/N^{1/(D-2)}$, and is much larger than $1/\sqrt{N}$
if $D>4$. This condition is precisely the condition for both equations
\radius\ and \entropy\ to be valid.

\newsec{\it Subtleties}

There are a number of subtleties one could raise to oppose the ideas 
put forward here. We shall mention only a couple of them. 

The first question is, if some special spin dependent interactions are 
important for understanding the large N limit of a black hole, 
what about other
spin dependent interactions? We have seen that the spin orbital coupling
$v^3\psi^2/r^8$ dominates the familiar force $v^4/r^7$ (in $D=11$), 
assuming that there is a correlation between spin and orbital motion
so that this force is not averaged out. This is because the typical
velocity in the large $N$ limit is very small, and the size of the 
black hole is fixed. Thus the smaller power in the velocity, the more
important a term is. For instance, the spin dependent force
$(l_p^9/R)v^2(\psi^2)^2/r^9$, if not vanishing, is larger than the one
we considered. However, if we use this term as the interaction energy
in \viria\ to determine
$R_s$, we will find that $R_s=N^{1/9}l_p$. The size of the cluster
blows up in the large N limit, not the canonical behavior of a boosted
transverse object. If the size does not behave canonically, it is
hard to demand the rest mass to behave canonically. In such a case,
the mean velocity is no longer suppressed by $1/N$, and we do not 
know whether the $v^2$ interaction continues to dominate the $v^3$
interaction. The same can be said of other types of spin dependent
interactions.

It is shown in \hm\ that at the transition point $N\sim S$ where
a black string becomes unstable and collapses to form a black hole,
one can use the two body interaction $v^4/r^7$ to determine the
size of the black hole, and of course the rest mass does not behave
canonically. It is easy to check that all super-partners of the
$v^4$ interaction are equally important in this regime. This fact
together our above discussion suggests that, in order to have the
canonical large N behavior, certain spin interaction as the one proposed
in the last section must be dominating, and the interacting gas must
be highly coherent such that other potentially more important spin
dependent interactions are actually switched off.

The second subtlety concerns the low velocity. We have assumed 
$v\sim TR_s$. For a fixed $R_s$, $T$ is very low in the large N
limit. It is easy to see that starting from a certain $N$ ($N=S$),
the mean velocity begins to become smaller than $1/R_s$ as set
by the uncertainty relation. One easy, but not constructive, way to
get around of this problem is to assume the strong holographic
principle hold \hm. In this case, the virtual size of the cluster is
not $R_s$, but some other scale much larger than $R_s$. According to
this strong holographic principle, one parton must occupy at least
a unit Planck cell, therefore the virtual transverse volume of the
cluster must be at least proportional to $N$. Another, we consider
more attractive resolution is to assume that the infinitely boosted
black hole is not a gas of partons, but a gas of threshold bound states
of certain size. Since our main equations \virial\ and \poten\ involve
the parameters $N$ and $R$ only through the combination $p_+=N/R$, our
calculations in the last section will go through if we replace partons
by threshold bound states of fixed size. It is possible to choose the
mass for these bound states such that the uncertainty relation is
not violated. Indeed, assuming the uncertainty relation be
saturated, using $v\sim TR_s$ and the relation between $T$ and $R_s$
as given in \radius, we find that the mass of the threshold bound state
is $N/(SR)$. To make up the total longitudinal momentum $p_+$, there
are precisely $S$ such bound states in the black hole. This is an 
indication that in the
large N limit, not all degrees of freedom, except only part of them,
are necessary for accounting for the black hole entropy.
One might wonder in such a case that whether it is still necessary to 
employ the spin dependent interactions of the last section. For instance, 
can one use the familiar force $v^4/r^7$ between two threshold bound 
states to obtain the desired result within the gas picture?
The answer to this question is no. To see this, recall that the corresponding
11D amplitude of two supergravitons is proportional to $p_{11}(1)p_{11}(2)$
\bfss, so the reduced amplitude is proportional to 
$(1/R)p_{11}(1)p_{11}(2)=N^2/(S^2R^3)$, thus the total interaction energy
would still be proportional to $N^2/R^3$ which does not lead to the
correct answer.

We do not exclude other possible resolution to the above puzzle.

\newsec{\it Fractal dimensions?}

It was suggested in the first paper of \bfks\ that under the condition
$N>>S$ a new phase appears in the super Yang-Mills theory (SYM) 
(for $d\le 3$). If one is to do thermodynamics on the dual torus 
$\tilde{T}^d$, one is not to expect to excite the usual momentum modes 
in the relevant SYM. These modes are necessarily longitudinal objects,
as pointed out in \hm, and therefore have nothing to do with
the infinitely boosted black hole. However, there still remains the
possibility of exciting some transverse modes, as the light-cone
energy of these modes is typically suppressed by a factor $1/N$, which
is of the same order as the temperature $T\sim 1/N$. It appears
that this scenario is in contradiction with what we have suggested, that
the black hole must be thought of as a highly coherent interacting
gas. Actually as far as we can see, there is no direct conflict. This is
because, so far all the known transverse objects are described
as some global objects living on $\tilde{T}^d$, for instance a transverse
membrane corresponds to a toron.

In this section we point out a curious form of formulas \entropy\ and
\radius\ which might lead us to some understanding in the context of SYM.
Let the size of the dual torus be $\Sigma$.
To see the dependence on the dual torus explicitly, we plug 
$L=l_p^3/(R\Sigma)$ into $G_D=l_p^9/L^d$. We now have
\eqn\curious{\eqalign{S &= N (N^{1\over d}\Sigma)^{{2d\over 7-d}}
(R/l_p^2)^{{3(d-3)\over 7-d}}T^{{9-d\over 7-d}},\cr
R_s &= l_p(N^{1\over d}\Sigma)^{{d\over 7-d}}(R/l_p^2)^{{d-1\over 7-d}}
T^{{1\over 7-d}}.}}
Now, the dependence on $N$ and $\Sigma$ combines into the effective
size $N^{1/d}\Sigma$, reminding us the mechanism employed by Maldacena
and Susskind to explain the fat black hole in the D-brane context
\ms. The physical picture for this is that all $N$ $Dd$-branes on
$\tilde{T}^d$ are connected to form a single large $Dd$-brane by switching
on Wilson lines in all directions. Since $\Sigma$ is enhanced, it is
possible to excite soft modes whose energy scales as $1/(N^{1/d}\Sigma)$.
And instead of $N^2$ species of light modes, there are now only $N$ 
species, this explains the extra factor $N$ in the formula for $S$.
There is also a nice interpretation for the factor $R/l_p^2$. As we
have seen in section 2, this is proportional to the parton coupling
constant.

The above interpretation is quite appealing, nevertheless
there is a big loop hole. The above argument is valid only in the
case when the excitations are dominantly momentum modes, and these are
not what we want to have. Even to excite momentum modes, the condition
is that the temperature $T$ must be greater than the inverse effective
size. Since the temperature goes as $1/N$, this is possible only when
$d=1$. For this case, a detailed calculation shows that the
condition $Sl_p>L$ must be satisfied, that is, the internal circle
is not too big. 

Taking equations \curious\ literally, we find that the effective
dimension is not $d$ but the fractal dimension $d_f=2d/(7-d)$. 
It is smaller than $d$ if $0<d<5$, and is equal to $d$ if $d=0,5$.
We know $d=5$ is a special case where multiple NS5-brane theory
is argued to be the correct matrix theory \seiberg. The strange
thing happens at $d=6$, here $d_f=12$ and is the only case where
$d_f>d$. This makes a fractal interpretation implausible. Precisely
starting from $T^6$, it is argued \ss\ that there is no simple 
matrix formulation of M-theory. We do not know whether there is a
connection between this fact and our observation.

\newsec{\it Discussion and conclusion}

Our discussion in this paper as well as the analysis of \hm\ strongly
suggests that an infinitely boosted black hole can be understood
on the basis of a strongly correlated gas of partons. Our treatment
is universal for all dimensions, except $D=4$ which requires a 
separate study. An interesting aspect of the coherent gas is that
the specific heat is always negative.
The main lesson learned here is that a D0-brane cluster
can have vastly different behavior in different kinetic regimes.
A black hole is certainly the regime where the cluster has the canonical 
behavior of a boosted transverse object. Another possible regime is
discussed in \aclm, where the size of the cluster is even bigger than 
the one discussed in \bfss. 
To understand more details of the working of the coherent 
gas for a black hole, much further work is required. In
particular, if one wishes to understand the matrix Schwarzschild
black hole from the standpoint of super Yang-Mills theories defined
on tori, new physics is to be invoked in the large N quantum field
theories. As pointed out in the previous section, the gas approach and
the SYM approach do not necessarily exclude each other. It might well
be that understanding gained in one approach will shed light 
on another approach. In any case, we expect that the black hole physics
is to teach us a lot about matrix theory, if matrix theory is
a viable model for M-theory.

\noindent{\bf Acknowledgments} 

We would like to thank S. Chaudhuri, J. Harvey, D. Kutasov, D. Minic,
T. Yoneya and especially E. Martinec for discussions. We are grateful
to G. Horowitz and E. Martinec for communicating their results before
publication.
This work was supported by DOE grant DE-FG02-90ER-40560 and NSF grant
PHY 91-23780.

\listrefs

\end